# Dielectric Detection of Single Nanoparticles Using a Microwave Resonator Integrated with a Nanopore

*Arda Secme[1,2,†,‡], Berk Kucukoglu[1,2,†,‡], Hadi S. Pisheh[1,2,†,‡], Yagmur Ceren Alatas[1,2], Uzay Tefek[1,2] H. Dilara Uslu[1,2], Batuhan E. Kaynak[1,2], Hashim Alhmoud[1,2], M. Selim Hanay[1,2,*]*

[1]Department of Mechanical Engineering, and

[2] UNAM – Institute of Materials Science and Nanotechnology,

Bilkent University, Ankara, 06800, Turkey

KEYWORDS: Nanoparticles, Microwave Sensors, Nanopore, Capacitive Detection, Resonant Sensing

ABSTRACT: The characterization of individual nanoparticles in a liquid constitutes a critical challenge for environmental, material, and biological sciences. To detect nanoparticles, electronic approaches are especially desirable owing to their compactness and lower costs. Indeed, for single-molecule and single-nanoparticle detection, resistive pulse sensing has advanced significantly during the last two decades. While resistive pulse sensing was widely used to obtain the geometric size information, impedimetric measurements to obtain dielectric signatures of nanoparticles have scarcely been reported. To explore this orthogonal sensing modality, we developed an impedimetric sensor based on a microwave resonator with a nanoscale sensing gap surrounding a nanopore. The approach of single nanoparticles near the sensing region and their translocation through the nanopore induced sudden changes in the impedance of the structure. The impedance changes in turn were picked up by the phase response of the microwave resonator. We worked with 100 nm and 50 nm polystyrene nanoparticles to observe single-particle events. Our current implementation was limited by the non-uniform electric field at the sensing region. The work provides a complementary sensing modality for nanoparticle characterization where the dielectric response, rather than the ionic current, determines the signal.

**INTRODUCTION**

Label-free detection of nanoparticles in liquid holds a key position for many applications in biology, material science, and environmental engineering. Nanoscale sensing techniques based on optical resonators,[1,2] suspended nanochannel resonators,[3] and resistive-pulse sensing across a nanopore[4] have all demonstrated the sizing of nanoparticles inside a liquid. Amongst these techniques, entirely electronic techniques, such as resistive-pulse sensing across a nanopore, offer the advantage of parallelism, low-cost, and portability. For important biological applications such as nucleotide and protein sequencing, nanopore-based techniques have already achieved commercialization[5] and proof-of-concept demonstrations,[6-12] respectively. Resistive-pulse sensing with nanopores[13] has also been used to detect single nanoparticles[4,14] and viruses.[15-17]

While resistive pulse sensing is an invaluable technique, it can be complemented by additional electronic measurements conducted on the same particle.[18] In resistive pulse sensing, the ionic current through a nanopore gets partially blocked by an insulating particle as the particle passes through the nanopore. Since the magnitude of the current blockage depends on the geometric size of the nanoparticles —and not on their material properties as long as the particles are insulating— particles of different compositions but similar sizes result in similar signals. Therefore, to obtain material-specific information in nanoparticle applications, complementary measurements are needed, such as the dielectric response of the nanoparticles. This way one can combine both geometric size and capacitive size measurements to obtain the dielectric permittivity of a material and hence perform permittivity-based material classification of nanoparticles[19] (Figure 1).

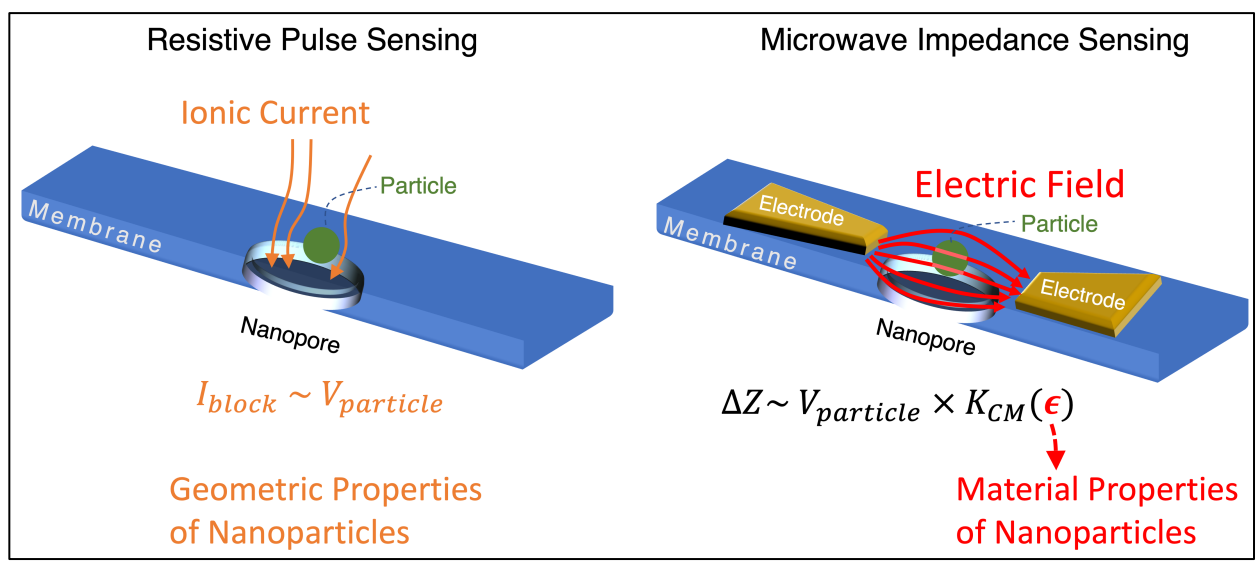

Figure 1. Motivation for microwave nanopore sensing. In typical resistive pulse sensing scenarios (left panel), the current across a nanopore is measured which gets blocked partially by a dielectric particle passing through the nanopore. Since the blockage current depends only on the size and shape of the particle, resistive-pulse sensing does not directly provide information about the internal properties. (Right Panel) At high frequencies, a nanoparticle can be probed through the impedance change it induces across two electrodes —which are part of a microwave resonator in this study (the microwave resonator is not shown here). This impedance change can be related to the real and imaginary parts of its dielectric permittivity, through the Clausius-Mossotti Factor $K_{CM}$.

In principle, the dielectric response of nanoparticles can be detected by a capacitive sensor with a nanoscale sensing region, at any given AC frequency. In practice, however, the operation of a capacitive sensor in liquid is hampered at low AC frequencies by Debye screening effects on the material interfaces, such as at the electrodes and on the surface of nanoparticles.[20] A potential remedy is to operate the electronic sensor at microwave frequencies to overcome such screening effects.[21, 22] Indeed, a scanning probe sensor with a microwave resonator (in air) has been shown to detect nanoparticles and obtain their dielectric signals.[23] Inside a liquid, scanning microwave microscopy has accomplished the imaging of internal bacteria structure,[24] and cells obstructed by a thin membrane.[25] Compared to probe-based microwave techniques, chip-based techniques offer direct integration capability with microfluidic platforms. The chip-based approach of interfacing a micropore with a microwave device was recently demonstrated in[22] for detecting 500 nm beads, which employed different measurement[26] and fabrication techniques compared to those reported here. In this work, we demonstrate that single nanoparticles of 50 and 100 nm diameter can be detected capacitively by on-chip microwave resonators fabricated in the proximity of a solid-state nanopore. In the following, we first describe the device paradigm and experimental system, followed by the detection of 100 nm polystyrene particles with this platform and their statistics.

EXPERIMENTS

The fabricated sensor is shown on Figure 2. We used a coplanar waveguide microwave resonator where the active sensing region is fabricated on a membrane, so that a nanopore can be drilled readily. The sensor was fabricated on a commercial wafer consisting of a 500 μm thick Si substrate and two additional material layers. The first layer was a 2.2 μm thick $SiO_2$ on top of Si, and the second (outermost) layer was a 220 nm thick $Si_3N_4$ layer. The wafer contained the $SiO_2$ and $Si_3N_4$ layers on both sides. During the fabrication (SI Section 1), we first processed the top side of the chip by performing electron beam lithography followed by metal deposition to define the narrow sensing region where two metal electrodes approach each other with a gap of approximately 600 nm. This step was followed by photolithography and metal deposition to define the larger features of the microwave sensor in the form of a coplanar waveguide. For both lithography steps, 10 nm Cr and 100 nm Au were deposited as the metallization layer.

After the process on the top surface was completed, we proceeded with membrane fabrication. A window was opened at the bottom of the wafer with anisotropic etching of $Si_3N_4$ to enable backside KOH wet etching. Si and $SiO_2$ were etched by KOH until the top $Si_3N_4$ layer was suspended which forms the thin membrane. Care was taken to align the backside window with the sensing region at the top as much as possible, this way the narrow sensing region at the top surface sits on the $Si_3N_4$ membrane. In the last step, Focused Ion Beam (FIB) was used to drill a hole on the membrane in between the sensing electrodes. In the final configuration, electrodes were aligned on the membrane (Fig. 2c,d) and a nanopore was opened between the electrodes, having typically 450 nm diameter, and 220 nm thickness determined by the thickness of $Si_3N_4$.

The fabricated chip was placed in a test setup inside a Faraday cage. The test setup contained a liquid reservoir with Phosphate Buffered Saline (PBS). The chip was partially submerged into the liquid reservoir at a distance that covers the bottom etching window completely (Figure 2a), while keeping the microwave readout port dry for external electronic access. A small liquid chamber called particle reservoir was attached to the top side of the chip and aligned with the sensing region so that the nanopore area is covered by the particle reservoir. Each reservoir was supplemented with a wire to induce electrokinetic flow of nanoparticles between the reservoirs and through the nanopore (the zeta potential is extrapolated to be between -50 to -60 mV, which implies electroosmotic flow). A sourcemeter (Keithley 2401) was used to measure the electronic characteristics between these two wires. One of the wires was placed inside the large reservoir where the sensor was also partially submerged, and the other wire was placed inside the particle reservoir (Figure S2). Current-voltage measurements between the two wires were conducted using the sourcemeter to verify that the two reservoirs were connected via the liquid which contains ions. The linearity of the I-V curves was established for the voltage ranges used in the experiments (Figure S2).

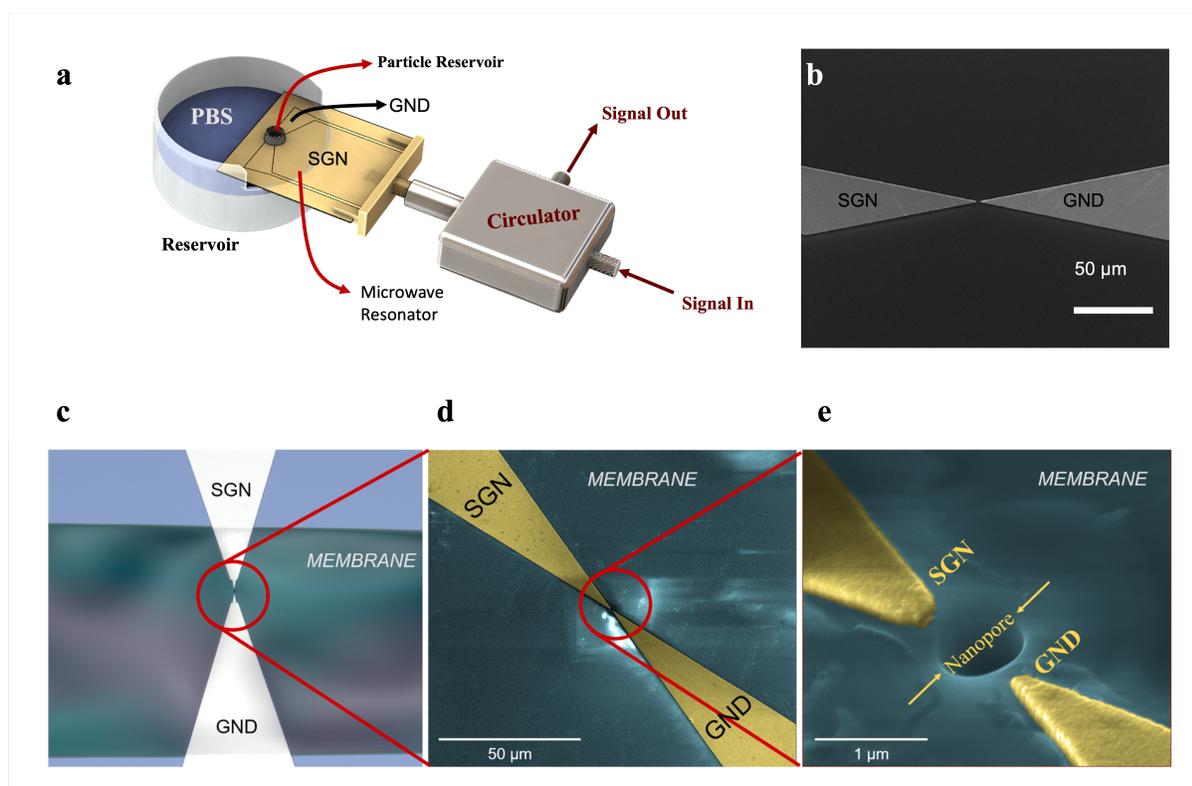

Figure 2. Experimental system and the device. (a) The setup consists of two reservoirs and a microwave resonator in the form of a coplanar waveguide resonator. The signal (SGN) and ground (GND) electrodes of the resonator converge together on a membrane. The particle reservoir is aligned with the membrane as well. (b) SEM image of the sensing region before the membrane and nanopore were fabricated. (c) Optical microscopy image of the membrane and the sensing region. (d, e) SEM images of the sensing region and the nanopore (tilted) at successive resolutions.

The microwave resonator used in the experiments was measured by a heterodyne mixing circuitry, also known as a microwave interferometer.[27-30] Owing to the frequency down-conversion of the signal, phase-sensitive detection using lock-in amplifiers in the 5 MHz range was employed. Further details of the measurement circuitry are described in SI Section 3. The first set of experiments was performed with polystyrene (PS) nanoparticles with 100 nm nominal diameter (Sigma Aldrich 43302). The original samples were diluted with PBS by 60-fold. Aliquots of 50 µL volume of the diluted solution were added to the top reservoir, located on the top surface of the chip. For the sensor used in the experiments, the electrodes were separated by 630 nm and the nanopore had a diameter of 450 nm. The resonance frequency of the device was at 6.62 GHz, and it was tracked with a sampling rate of 50 kSa/s and lock-in time constant of 50 µs which are sufficient to resolve single events that occur within a duration of several milliseconds.

In the experiments, before introducing the nanoparticles into the liquid chamber, control experiments were conducted first, where a PBS solution with no nanoparticles are introduced and 0.5 V was applied to induce electrokinetic motion. The resulting data trace does not contain any sharp features (Figure 3a, blue trace). By contrast when the 100 nm polystyrene solution was introduced, frequent spike-like transitions were observed typically with several millisecond durations indicative of single-particle translocation events (Figure 3a orange trace, Figure 3b). Ramping up the electrokinetic voltage increased how often these spikes occurred (Figure S4). Since spikes were absent in the control experiments, they were designated to originate from single nanoparticles approaching the sensing region.

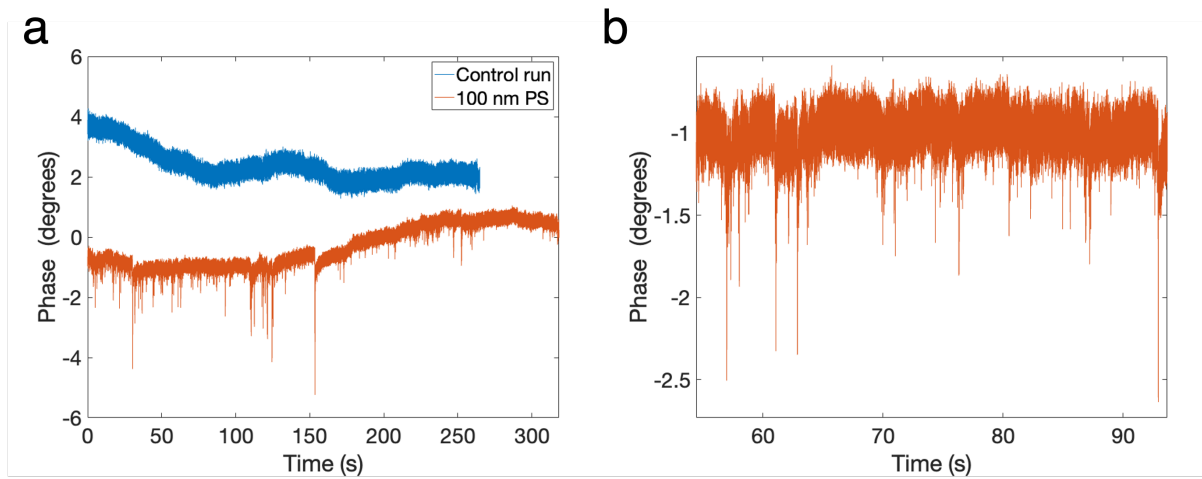

Figure 3. Detecting 100 nm polystyrene particles with microwave nanopore setup. (a) Control run (blue) and the actual run (orange) with spike-like events. The control run is shown with a 4º phase offset in the y-axis so that the data traces do not overlap (b) Close-up view of some of the events in part (a).

The histogram for the magnitude of the phase shifts and the event durations were obtained from the data (Figure 4). The histograms were obtained by first filtering the time-series data with a low-pass filter to remove the excess noise (Section S6). The prominence and width values of the spike-like events were determined by the built-in MATLAB functions. A threshold for prominence value was set so that events smaller than the half of the peak-to-peak noise level of control runs were discarded. With this stringent threshold setting, the spike detection algorithm did not register any events in the control runs, as expected. For this reason, all the events detected above the threshold were interpreted to originate from nanoparticles migrating near the active sensing region.

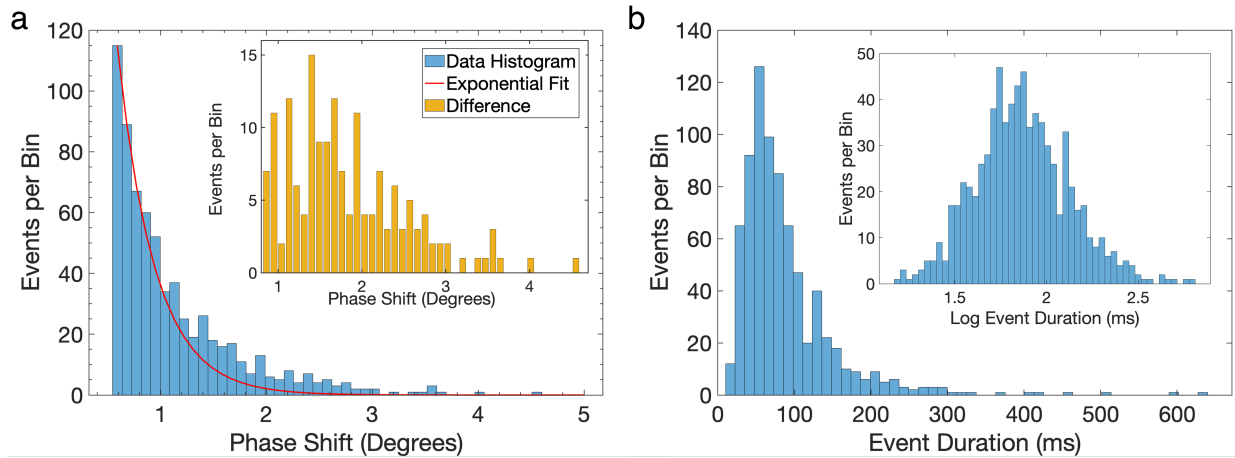

Figure 4. Statistics of 100 nm PS events. (a) Histogram of the frequency shifts are shown as blue bars, whereas the red curve is an exponential fit adjusted with respect to the decay rates of the first few bins. The inset shows the difference between the data histogram and the exponential fit. For clarity, inset event values are rounded up to the nearest integer and reported for events with phase shift > 0.8°. (b) Histogram for the duration of the events. Inset: the same information is shown with a logarithmic x-axis.

One of the salient features of the phase shift histogram (Figure 4a) is the large number of events at low values and a decaying tail in the distribution. At first, this tail may appear to originate from the noise: however, the stringent criterion for spike detection detailed above, and the absence of any events during control experiments, indicate otherwise. To further understand this behavior, we fit an exponential curve based on the first few bins of the histogram. After the fitting curve decays to almost zero at large phase shift values, the histogram values remain significantly larger than what is expected by the exponential decay, indicating a secondary population of events. This population —the difference between the original histogram and the exponential fitting— is shown on the inset of Figure 4a. The histogram for the subpopulation has a peak around 1.4° phase shift which is close to the expected size of the phase shifts for 100 nm PS particles as discussed in the next section.

**Capacitive changes induced by nanoparticles**

The measured phase shifts in Figure 4 can be related to the impedance change induced by the nanoparticles with the real and imaginary parts of the dielectric constant determining the real and imaginary parts of the response, respectively. To obtain the real and imaginary parts, both the amplitude and phase changes of the resonator can be tracked. Within the existing system, only the real part, i.e. the capacitive contribution, of the nanoparticle is accessible (SI Sections S9 and S10), so we focus the discussion on the capacitance change induced by a particle of volume $V_{particle}$:[28]

$$\Delta C = V_{particle} \times 3 K_{CM} \times \epsilon_m \times \frac{|E_{rms}(r_{particle})|^2}{U_{rms}^2} \quad ...(1)$$

where $\epsilon_m$ is the permittivity of the medium (e.g. aqueous solution), $E_{rms}(r_{particle})$ is the root-mean squared (rms) Electrical field at the particle location, and $U_{rms}$ is the rms value of the voltage difference. Finally, $K_{CM}$ stands for the Clausius-Mossotti Factor which is a function of the permittivity values of the particle ($\epsilon_p$) and the medium ($\epsilon_m$):

$$K_{CM} = \frac{\epsilon_p - \epsilon_m}{\epsilon_p + 2\epsilon_m} \quad ...(2)$$

Once the expected capacitance is thus obtained by equation (1), it can be first related to the change in the resonance frequency of the sensor, which in turn can be translated into phase shifts by using the phase versus frequency curve of the sensor which is experimentally obtained before the experiments (Supplementary Section S5). This way the capacitance change induced by nanoparticles can be obtained. We note that, the capacitance change depends not only on the properties of the nanoparticle, but also on the sensor, such as the distance between the sensing electrodes which determines the local electric field, $E_{rms}(r_{particle})$ at the nanoparticle location.

In the measurements, after subtracting the exponential decay as in Figure 4a, the peak value for phase shift distribution is 1.4°. The slope of the phase versus frequency of the detection circuit is 64.5° / MHz. Thus, a typical nanoparticle event of 1.4° translates into a 21.7 kHz shift in resonance frequency, for a 6.6 GHz resonator. Converting this frequency shift into a fractional capacitance change ( $\left|\frac{\Delta C}{C}\right| = 2\frac{\Delta f}{f}$), we obtain $6.6 \times 10^{-6}$ fractional change in capacitance induced by a nanoparticle. This value can be compared to the expected capacitance change from equation (1). A 100 nm polystyrene particle (with a relative permittivity of 3.0) passing through an interelectrode distance of 630 nm is expected to induce a $5.4 \times 10^{-6}$ fractional change in capacitance, which is close to the measured value, $6.6 \times 10^{-6}$. Therefore, the size of the observed spikes in this region matches the expected effect of a nanoparticle passing through the nanopore.

To show the broader applicability of the technique, we also tested 50 nm size polystyrene particles and Felocell vaccine (Zoetis). We used different resonators for each species, due to the fragility of the devices (device parameters are listed on SI Section S11). The 50 nm PS particles were diluted by 100-fold and 25 $\mu$L of solution was introduced to the sample reservoir. Felocell vaccine contains virion particles of varying sizes (Feline Rhinotracheitis: 150–200 nm, Calicivirus: 35–39 nm, Panleukopenia: 18–22 nm). The vaccine was in lyophilized form, so it was reconstituted with PBS and filtered with a 220 nm pore sized filter. After the mixture was pipetted to forestall clustering, 50 µL of viruses in PBS was added to the particle reservoir. Data analysis for both cases were conducted as described before. In both cases, control runs under the same experimental analytes resulted again in data traces with no visible spikes. On the other hand, test runs with the analytes resulted in spike-shaped events and phase shift histograms similar to the 100 nm PS histogram before (Figure 5): a decaying component at lower values and a subpopulation of events at larger frequencies.

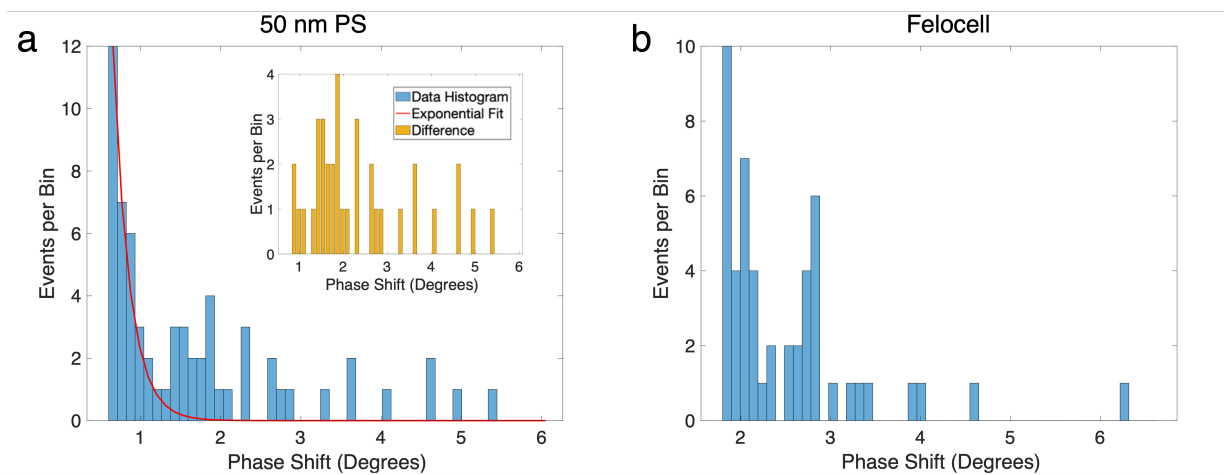

Figure 5. 50 nm PS and Felocell vaccine experiments. (a) Histogram of the frequency shifts for 50 nm polystyrene particles. The red curve is an exponential fit adjusted with respect to the decay rates of the first few bins. The inset shows the difference between the data histogram and the exponential fit. For clarity, inset event values are rounded up to the nearest integer and reported for events with phase shift > 0.6º (b) Histogram for Felocell virions.

**DISCUSSION**

The broad nature of the phase shift histogram (Figure 4a) requires further consideration since its large dispersion creates an impediment to a practical sensing scenario. Part of the dispersion may be related to the inherent polydispersity of the sample. A 10 nm standard deviation is reported for 100 nm particles, which translates into a 33% coefficient of variation for particle volume. While this variation is significant, it cannot by itself account for the broadness of the histograms. Here, we consider two additional factors that can contribute to signal broadening: the positional dependency of the sensor response in the nanopore region and particles approaching but not translating through the nanopore.

In the device design, the electrode shapes were chosen to be almost triangular to create sharp tips for increasing the electric field intensity (Figure 1e). After nanofabrication, the width of the tip corners appeared to have a finite size of approximately 220 nm from SEM measurements. Since this width is smaller than the diameter of the nanopore (450 nm), the electric field in the sensing region is not uniform due to the fringing field effects. Since the signal generated by a passing particle is proportional to the square of the electrical field, the response of the sensor depends strongly on where a nanoparticle passes through the nanopore. For instance, the center of the nanopore is aligned with the centers of the electrodes where the electric field is maximum: as a result, a nanoparticle passing through the nanopore center will generate a large response. However, the signal magnitude will be smaller for a nanoparticle passing near the edge of the nanopore. We conducted 3D simulations to obtain the overall trend under these conditions (Figure 6, SI Section S7). In this case, the sensing region with the electrodes, membrane and nanopore were modeled inside water to obtain the electric potential and field (Figure 6, inset). Using the extracted electric field magnitudes, a Monte Carlo simulation was conducted to calculate the response of the nanoparticles, assuming a uniform distribution through the nanopore (excluding edges with a margin of 50 nm so that the nanoparticles can fit). The response of each particle was calculated as the integral of the square of the electrical field for the volume of each 100 nm particle (at the vertical position where their response is maximal). The result qualitatively captures that a large portion of the signals occurs at lower values, which decay out with a tail at high values. Thus, the electrical inhomogeneity and positional dependency of the sensor partially explain the decaying trend in the experimental histogram.

A second potential factor for the broad distribution of the events is that a nanoparticle can still induce a signal even if it does not translocate through the nanopore: it is sufficient that they migrate to a region with large enough electric field strength. The field near the nanopore region is large enough to result in a nanoparticle signal above the noise floor, but not large enough to induce full response that would have occurred if the particle passed through the center of the nanopore. Such incidents may further contribute to the number of events giving rise to signals at the lower values of the histogram. By contrast, resistive pulse sensing technique requires the particle to be present inside the nanopore, so that a blockage current can be generated. This requirement is an advantage for obtaining uniform signals in resistive pulse sensing. However, the proximal sensing capability of capacitive sensors can offer an advantage to study the dynamics of nanoparticles and other analytes as they approach the nanopore.

We note that, even with a uniform electric field, the capability for material classification requires the attainment of very high signal-to-noise ratio in microwave sensing.[31] This requirement is due to the fact that the Clausius-Mossotti factor depends weakly on the particle permittivity, when the permittivity of the medium (e.g. water) is large. Apart from attaining large signal-to-noise ratio, changing the type of the medium could be another solution for environmental applications.

In this work, we operated the system only at a single frequency which provides only parameters at that frequency. To obtain multifrequency data, either a broadband measurement, or multiplexed narrowband detection[30, 32] could be employed in future work. In the analysis of nanoparticle data (SI Section S5), only the capacitance change of the bare particle was considered; the ionic shell surrounding the particle was ignore for the sake of simplicity. Due to the fragility of the nanomembrane devices, we had to use different devices in different runs. Due to the non-uniformity of the electrical field (e.g., Figure 6 inset), device-to-device variations especially in the active region, and the uncertainty in the sensitivity analysis (SI Section S8), no attempt was made to compare the different sample runs with each other at this stage. While microwave interferometer topology increases the sensitivity, it is not entirely clear if this topology also introduced extra noise on the measurements.

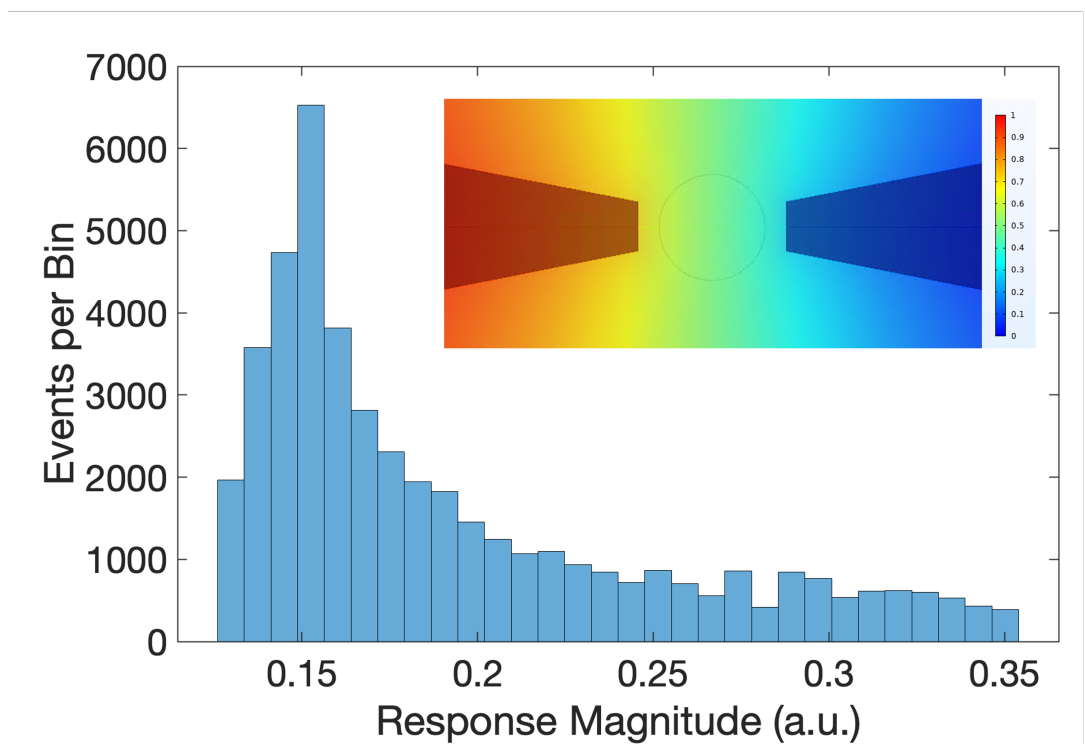

Figure 6: Monte Carlo simulation results for signal size distribution. The inset shows the electric potential distribution in the sensing region for the simulations, with the outline of the nanopore visible between the two electrodes. The colormap shows voltage levels.

The stage of development for microwave nanoparticle sensors can be compared to the early days of nanomechanical mass spectrometry,[33, 34] as well as impedance cytometry,[35] where position dependency of the analytes conspired to broaden the signal values. Later on, the use of multiple modes in nanomechanics,[36, 37] and multielectrode approaches in impedance cytometry[38-40] were invented to factor out positional effects. Similarly, the results obtained here motivates the development of new device designs and measurement techniques to eliminate positional dependency of nanoparticle signals for capacitive sensing. Beyond single nanoparticle sensing reported here, microwave capacitive sensing can be applied to dynamically characterize ion distributions inside nanoscale channels to provide a novel characterization technique for nanofluidics.[41-43]

**CONCLUSION**

In conclusion we demonstrated proof-of-concept experiments for capacitive sensing of nanoparticles as they migrate close to a nanoscale sensing region and translocate through a nanopore. The measurements are conducted by tracking the phase response of a microwave resonator. As a particle modulates the capacitance of the microwave resonator, the resonator phase gets modulated as well, which is detected by a custom measurement circuitry. The measured phase shift can be related first to the capacitance change induced by the particle, and then to its electrical size. Capacitive sensing is critical to provide both material-dependent information in electronic sensing, as well as understanding the transient dynamics of analytes as they approach nanopore region. With the proof-of-principle demonstration of on-chip microwave sensing, the stage is ripe for dielectric sizing and characterization of nanoparticles capacitively in liquid.

ASSOCIATED CONTENT

The following files are available free of charge.

Figures describing the experimental system and auxiliary data figures (PDF)

AUTHOR INFORMATION

**Corresponding Author**

* selimhanay@bilkent.edu.tr

**Present Addresses**

†A. Secme: California Institute of Technology, Pasadena, CA

B. Kucukoglu: EPFL, Lausanne, Switzerland

H. S. Pisheh: MDPI, Erlangen, Germany

**Author Contributions**

The manuscript was written through contributions of all authors. All authors have given approval to the final version of the manuscript. ‡These authors contributed equally. HSP fabricated the devices; AS, HDU and BK conducted the sensing experiments; BK, AS and MSH performed the data analysis; AS, BK and MSH wrote the manuscript.

**Funding Sources**

This project has received funding from the European Research Council (ERC) under the European Union's Horizon 2020 research and innovation programme (grant agreement n° 758769).

ACKNOWLEDGMENT

The authors thank Mehmet Kelleci, Burak Sari and Mohammed Alkhaled for useful discussions.

REFERENCES

(1) Vollmer, F.; Arnold, S.; Keng, D. Single virus detection from the reactive shift of a whispering-gallery mode. *Proc. Natl. Acad. Sci. U.S.A.* **2008**, *105* (52), 20701-20704.

(2) He, L.; Özdemir, Ş. K.; Zhu, J.; Kim, W.; Yang, L. Detecting single viruses and nanoparticles using whispering gallery microlasers. *Nat. Nanotechnol.* **2011**, *6* (7), 428-432.

(3) Olcum, S.; Cermak, N.; Wasserman, S. C.; Christine, K. S.; Atsumi, H.; Payer, K. R.; Shen, W.; Lee, J.; Belcher, A. M.; Bhatia, S. N. Weighing nanoparticles in solution at the attogram scale. *Proc. Natl. Acad. Sci. U.S.A.* **2014**, *111* (4), 1310-1315.

(4) Fraikin, J.-L.; Teesalu, T.; McKenney, C. M.; Ruoslahti, E.; Cleland, A. N. A high-throughput label-free nanoparticle analyser. *Nature Nanotechnology* **2011**, *6* (5), 308-313.

(5) Brown, C. G.; Clarke, J. Nanopore development at Oxford nanopore. *Nat. Biotechnol.* **2016**, *34* (8), 810-811.

(6) Piguet, F.; Ouldali, H.; Pastoriza-Gallego, M.; Manivet, P.; Pelta, J.; Oukhaled, A. Identification of single amino acid differences in uniformly charged homopolymeric peptides with aerolysin nanopore. *Nat. Commun.* **2018**, *9* (1), 966.

(7) Wang, R.; Gilboa, T.; Song, J.; Huttner, D.; Grinstaff, M. W.; Meller, A. Single-molecule discrimination of labeled DNAs and polypeptides using photoluminescent-free TiO2 nanopores. *ACS Nano* **2018**, *12* (11), 11648-11656.

(8) Restrepo-Pérez, L.; Wong, C. H.; Maglia, G.; Dekker, C.; Joo, C. Label-free detection of post-translational modifications with a nanopore. *Nano Lett.* **2019**, *19* (11), 7957-7964.

(9) Ouldali, H.; Sarthak, K.; Ensslen, T.; Piguet, F.; Manivet, P.; Pelta, J.; Behrends, J. C.; Aksimentiev, A.; Oukhaled, A. Electrical recognition of the twenty proteinogenic amino acids using an aerolysin nanopore. *Nat. Biotechnol.* **2020**, *38* (2), 176-181.

(10) Brinkerhoff, H.; Kang, A. S.; Liu, J.; Aksimentiev, A.; Dekker, C. Multiple rereads of single proteins at single–amino acid resolution using nanopores. *Science* **2021**, *374* (6574), 1509-1513.

(11) Afshar Bakshloo, M.; Kasianowicz, J. J.; Pastoriza-Gallego, M.; Mathé, J.; Daniel, R.; Piguet, F.; Oukhaled, A. Nanopore-based protein identification. *J. Am. Chem. Soc.* **2022**, *144* (6), 2716-2725.

(12) Ying, Y.-L.; Hu, Z.-L.; Zhang, S.; Qing, Y.; Fragasso, A.; Maglia, G.; Meller, A.; Bayley, H.; Dekker, C.; Long, Y.-T. Nanopore-based technologies beyond DNA sequencing. *Nat. Nanotechnol.* **2022**, *17* (11), 1136-1146.

(13) Haq, M. R.; Lee, B. J.; Lee, J. Solid-State Nanopore for Molecular Detection. *Int. J. Precis. Eng.* **2021**, 1-26.

(14) Tsutsui, M.; Yamazaki, T.; Tatematsu, K.; Yokota, K.; Esaki, Y.; Kubo, Y.; Deguchi, H.; Arima, A.; Kuroda, S. i.; Kawai, T. High-throughput single nanoparticle detection using a feed-through channel-integrated nanopore. *Nanoscale* **2019**, *11* (43), 20475-20484.

(15) Yang, L.; Yamamoto, T. Quantification of virus particles using nanopore-based resistive-pulse sensing techniques. *Front. Microbiol.* **2016**, *7*, 1500.

(16) Uram, J. D.; Ke, K.; Hunt, A. J.; Mayer, M. Submicrometer pore-based characterization and quantification of antibody–virus interactions. *Small* **2006**, *2* (8-9), 967-972.

(17) Zhou, K.; Li, L.; Tan, Z.; Zlotnick, A.; Jacobson, S. C. Characterization of hepatitis B virus capsids by resistive-pulse sensing. *J. Am. Chem. Soc.* **2011**, *133* (6), 1618-1621.

(18) Sheng, Q.; Wang, X.; Xie, Y.; Wang, C.; Xue, J. A capacitive-pulse model for nanoparticle sensing by single conical nanochannels. *Nanoscale* **2016**, *8* (3), 1565-1571.

(19) Tefek, U.; Sari, B.; Alhmoud, H. Z.; Hanay, M. S. Permittivity-Based Microparticle Classification by the Integration of Impedance Cytometry and Microwave Resonators. *Advanced Materials* **2023**, 2304072.

(20) Bazant, M. Z.; Thornton, K.; Ajdari, A. Diffuse-charge dynamics in electrochemical systems. *Physical review E* **2004**, *70* (2), 021506.

(21) Laborde, C.; Pittino, F.; Verhoeven, H.; Lemay, S.; Selmi, L.; Jongsma, M.; Widdershoven, F. Real-time imaging of microparticles and living cells with CMOS nanocapacitor arrays. *Nat. Nanotechnol.* **2015**, *10* (9), 791-795.

(22) Bhat, A.; Gwozdz, P. V.; Seshadri, A.; Hoeft, M.; Blick, R. H. Tank circuit for ultrafast single-particle detection in micropores. *Physical review letters* **2018**, *121* (7), 078102.

(23) Fumagalli, L.; Esteban-Ferrer, D.; Cuervo, A.; Carrascosa, J. L.; Gomila, G. Label-free identification of single dielectric nanoparticles and viruses with ultraweak polarization forces. *Nat. Mater.* **2012**, *11* (9), 808-816.

(24) Biagi, M. C.; Fabregas, R.; Gramse, G.; Van Der Hofstadt, M.; Juarez, A.; Kienberger, F.; Fumagalli, L.; Gomila, G. Nanoscale electric permittivity of single bacterial cells at gigahertz frequencies by scanning microwave microscopy. *ACS Nano* **2016**, *10* (1), 280-288.

(25) Tselev, A.; Velmurugan, J.; Ievlev, A. V.; Kalinin, S. V.; Kolmakov, A. Seeing through walls at the nanoscale: Microwave microscopy of enclosed objects and processes in liquids. *ACS Nano* **2016**, *10* (3), 3562-3570.

(26) Secme, A.; Tefek, U.; Sari, B.; Pisheh, H. S.; Uslu, H. D.; Çalışkan, Ö. A.; Kucukoglu, B.; Erdogan, R. T.; Alhmoud, H.; Sahin, O. High-Resolution Dielectric Characterization of Single Cells and Microparticles Using Integrated Microfluidic Microwave Sensors. *IEEE Sensors Journal* **2023**, *23* (7), 6517-6529.

(27) Nikolic-Jaric, M.; Romanuik, S.; Ferrier, G.; Bridges, G.; Butler, M.; Sunley, K.; Thomson, D.; Freeman, M. Microwave frequency sensor for detection of biological cells in microfluidic channels. *Biomicrofluidics* **2009**, *3* (3), 034103.

(28) Ferrier, G. A.; Romanuik, S. F.; Thomson, D. J.; Bridges, G. E.; Freeman, M. R. A microwave interferometric system for simultaneous actuation and detection of single biological cells. *Lab Chip* **2009**, *9* (23), 3406-3412.

(29) Afshar, S.; Salimi, E.; Braasch, K.; Butler, M.; Thomson, D. J.; Bridges, G. E. Multi-frequency DEP cytometer employing a microwave sensor for dielectric analysis of single cells. *IEEE Transactions on Microwave Theory and Techniques* **2016**, *64* (3), 991-998.

(30) Kelleci, M.; Aydogmus, H.; Aslanbas, L.; Erbil, S. O.; Hanay, M. S. Towards microwave imaging of cells. *Lab on a Chip* **2018**, *18* (3), 463-472.

(31) Sari, B.; Tefek, U.; Hanay, M. S. Classification of Dielectric Microparticles by Microwave Impedance Cytometry. *bioRxiv* **2022**.

(32) Chien, J.-C.; Ameri, A.; Yeh, E.-C.; Killilea, A. N.; Anwar, M.; Niknejad, A. M. A high-throughput flow cytometry-on-a-CMOS platform for single-cell dielectric spectroscopy at microwave frequencies. *Lab on a Chip* **2018**, *18* (14), 2065-2076.

(33) Naik, A. K.; Hanay, M.; Hiebert, W.; Feng, X.; Roukes, M. L. Towards single-molecule nanomechanical mass spectrometry. *Nature nanotechnology* **2009**, *4* (7), 445-450.

(34) Chaste, J.; Eichler, A.; Moser, J.; Ceballos, G.; Rurali, R.; Bachtold, A. A nanomechanical mass sensor with yoctogram resolution. *Nature nanotechnology* **2012**, *7* (5), 301-304.

(35) Daguerre, H.; Solsona, M.; Cottet, J.; Gauthier, M.; Renaud, P.; Bolopion, A. Positional dependence of particles and cells in microfluidic electrical impedance flow cytometry: Origin, challenges and opportunities. *Lab Chip* **2020**, *20* (20), 3665-3689.

(36) Dohn, S.; Svendsen, W.; Boisen, A.; Hansen, O. Mass and position determination of attached particles on cantilever based mass sensors. *Rev. Sci. Instrum.* **2007**, *78* (10), 103303. DOI: 10.1063/1.2804074.

(37) Hanay, M. S.; Kelber, S.; Naik, A. K.; Chi, D.; Hentz, S.; Bullard, E. C.; Colinet, E.; Duraffourg, L.; Roukes, M. L. Single-protein nanomechanical mass spectrometry in real time. *Nat. Nanotechnol.* **2012**, *7* (9), 602-608. DOI: 10.1038/nnano.2012.119.

(38) Spencer, D.; Caselli, F.; Bisegna, P.; Morgan, H. High accuracy particle analysis using sheathless microfluidic impedance cytometry. *Lab Chip* **2016**, *16* (13), 2467-2473.

(39) De Ninno, A.; Errico, V.; Bertani, F. R.; Businaro, L.; Bisegna, P.; Caselli, F. Coplanar electrode microfluidic chip enabling accurate sheathless impedance cytometry. *Lab on a Chip* **2017**, *17* (6), 1158-1166.

(40) Spencer, D. C.; Paton, T. F.; Mulroney, K. T.; Inglis, T. J.; Sutton, J. M.; Morgan, H. A fast impedance-based antimicrobial susceptibility test. *Nat. Commun.* **2020**, *11* (1), 5328.

(41) Robin, P.; Bocquet, L. Nanofluidics at the crossroads. *arXiv preprint arXiv:2301.08829* **2023**.

(42) Robin, P.; Emmerich, T.; Ismail, A.; Niguès, A.; You, Y.; Nam, G.-H.; Keerthi, A.; Siria, A.; Geim, A.; Radha, B. Long-term memory and synapse-like dynamics in two-dimensional nanofluidic channels. *Science* **2023**, *379* (6628), 161-167.

(43) Xiong, T.; Li, C.; He, X.; Xie, B.; Zong, J.; Jiang, Y.; Ma, W.; Wu, F.; Fei, J.; Yu, P. Neuromorphic functions with a polyelectrolyte-confined fluidic memristor. *Science* **2023**, *379* (6628), 156-161.